\documentclass[modern]{aastex62}

\usepackage{graphicx}
\usepackage{amsmath}
\usepackage{longtable}
\usepackage{color}
\usepackage{enumerate}
\usepackage{graphicx,epstopdf}
\usepackage{subfigure}

\graphicspath{{./}{figures/}}

\shorttitle{BH candidates by LAMOST and ASAS-SN}
\shortauthors{Zheng et al.}

\def\eff{\rm eff}
\def\dex{\rm dex}
\def\K{\rm K}

\def\S/N{\rm S/N}
\def\km/s{\rm km~s^{-1}}

\begin{document}

\title{Searching for Black Hole Candidates by LAMOST and ASAS-SN}

\correspondingauthor{Wei-Min Gu}
\email{guwm@xmu.edu.cn}

\author{Ling-Lin Zheng}
\affiliation{Department of Astronomy, Xiamen University,
Xiamen, Fujian 361005, P. R. China}
\author{Wei-Min Gu}
\affiliation{Department of Astronomy, Xiamen University,
Xiamen, Fujian 361005, P. R. China}
\author{Tuan Yi}
\affiliation{Department of Astronomy, Xiamen University,
Xiamen, Fujian 361005, P. R. China}
\author{Jin-Bo Fu}
\affiliation{Department of Astronomy, Xiamen University,
Xiamen, Fujian 361005, P. R. China}
\author{Hui-Jun Mu}
\affiliation{Department of Astronomy, Xiamen University,
Xiamen, Fujian 361005, P. R. China}
\author{Fan Yang}
\affiliation{National Astronomical Observatories, Chinese
Academy of Sciences, Beijing 100012, P. R. China}
\affiliation{Infrared Processing and Analysis Center,
California Institute of Technology, Pasadena, CA 91125, USA}
\author{Song Wang}
\affiliation{National Astronomical Observatories, Chinese
Academy of Sciences, Beijing 100012, P. R. China}
\author{Zhong-Rui Bai}
\affiliation{National Astronomical Observatories, Chinese
Academy of Sciences, Beijing 100012, P. R. China}
\author{Hao Sou}
\affiliation{Department of Astronomy, Xiamen University,
Xiamen, Fujian 361005, P. R. China}
\author{Yu Bai}
\affiliation{National Astronomical Observatories, Chinese
Academy of Sciences, Beijing 100012, P. R. China}
\author{Yi-Ze Dong}
\affiliation{Department of Astronomy, Xiamen University,
Xiamen, Fujian 361005, P. R. China}
\affiliation{Department of Physics, University of California, Davis, CA 95616, USA}
\author{Hao-Tong Zhang}
\affiliation{National Astronomical Observatories, Chinese
Academy of Sciences, Beijing 100012, P. R. China}
\author{Ya-Juan Lei}
\affiliation{National Astronomical Observatories, Chinese
Academy of Sciences, Beijing 100012, P. R. China}
\author{Junfeng Wang}
\affiliation{Department of Astronomy, Xiamen University,
Xiamen, Fujian 361005, P. R. China}
\author{Jianfeng Wu}
\affiliation{Department of Astronomy, Xiamen University,
Xiamen, Fujian 361005, P. R. China}
\author{Jifeng Liu}
\affiliation{National Astronomical Observatories, Chinese
Academy of Sciences, Beijing 100012, P. R. China}
\affiliation{College of Astronomy and Space Sciences,
University of Chinese Academy of Sciences, Beijing 100049,
P. R. China}

\begin{abstract}
Most dynamically confirmed stellar-mass black holes
and the candidates were originally selected from
X-ray outbursts. In the present work, we search for
black hole candidates in the LAMOST
survey by using the spectra along with photometry
from the ASAS-SN survey, where the orbital period
of the binary may be revealed by the periodic light
curve, such as the ellipsoidal modulation type.
Our sample consists of 9 binaries,
where each source contains a giant star with large
radial velocity variation
($\Delta V_{\rm R} \ga 70~{\rm km~s^{-1}}$)
and periods known from light curves.
We focus on the 9 sources with long periods
($T_{\rm ph} > 5$~days)
and evaluate the mass $M_2$ of the optically invisible
companion. Since the observed $\Delta V_{\rm R}$ from
only a few repeating spectroscopic observations is a
lower limit of the real amplitude, the real mass $M_2$
can be significantly higher than the current evaluation.
It is likely an efficient method to place constraints
on $M_2$ by combining $\Delta V_{\rm R}$ from LAMOST
and $T_{\rm ph}$ from ASAS-SN, particularly by the
ongoing LAMOST Medium Resolution Survey.
\end{abstract}


\section{Introduction}\label{sec1}

It is well-known that three types of compact objects
in the Universe are white dwarfs, neutron stars, and
black holes (BHs). Since an isolated BH does not
produce electromagnetic radiation, most confirmed
stellar-mass BHs and candidates were found in binaries \citep{RM2006}. For a binary system composed of a BH
and an optically visible star filling its Roche lobe,
the matter from the star can be accreted by
the BH through the inner Lagrange point.
In such cases, an accretion disk is formed and X-ray
emission is produced from the disk.
Thus, such a BH binary system can be detected by X-ray
telescopes. However, the number of confirmed BHs and
BH candidates found by this method is less than
a hundred \citep{Corral-Santana2016}, which is far
below the number of BHs that are thought to exist
in our Galaxy \citep[e.g.,][]{Brown1994}.

New methods are required to search for more BH
candidates. For binaries with unknown orbital periods, \citet{Gu2019} proposed a method to search for BH
candidates from optical observations. The method is
based on the assumption that the radius $R_1$ of the
optically visible star is no more than the corresponding
Roche-lobe radius $R_{\rm L1}$. On the other hand,
once the orbital period $P_{\rm orb}$ can be derived
(such as being revealed by the periodic light curves),
we can obtain the well-known mass function (refer to
Equation~(\ref{E5}) in Section~\ref{sec3.2}) and
therefore place better constraints on the optically
invisible companion. In a BH binary, if the ratio
$R_1/R_{\rm L1}$ is not far below unity, the
companion may be pulled into a waterdrop shape due
to the strong gravity of the BH. The deformed star
will present a periodic light curve with the ellipsoidal
modulation \citep{Morris1985}. Thus, the light curve
may reveal the orbital period $P_{\rm orb}$ of the
system and is helpful to the constraints of $M_2$.

LAMOST (Large Sky Area Multi-Object Fiber Spectroscopic
Telescope) provides nearly 10 million stellar spectra
in the Data Release 6 and about 480 thousand
low resolution stellar spectra in the Data Release 7.
Furthermore, it has radial velocity to a precision
of better than 5~${\rm km~s^{-1}}$ \citep{Deng2012}.
We can derive many key parameters ($T_{\rm eff}$,
$\log g$, and [Fe/H]) and heliocentric radial velocity
$V_{\rm R}$ from the spectra \citep{Zong2018}.
In addition, ASAS-SN monitors the entire visible sky to a
depth of V$\lesssim17 $ {$\rm mag$} for bright supernovae
and other transients. There are nearly 430 thousand
variable stars in the catalog \citep{Jayasinghe2019}.

The aim of this paper is to introduce the method to
search for black hole candidates by combining the
LAMOST spectra and the ASAS-SN photometry. We will
introduce the data selection from LAMOST in
Section~\ref{sec2}. The analyses and results of our
sample are shown in Section~\ref{sec3}. Conclusions
and discussion are presented in Section~\ref{sec4}.

\section{Data selection}\label{sec2}

The present work focuses on binaries with a giant star.
For a giant star, the variation of radial velocity in
the same night is usually negligible due to its large
size, and therefore its orbital period is relatively
long. We select a sample of binaries containing a giant
star from LAMOST Data Release 6 and LAMOST Data Release 7
with the following criteria:

$\begin{cases}
{\rm S/N_{(g)}} > 10~($signal-to-noise in the g band$), \\
3800~\K < {\it T}_{\eff} < 5300~\K  , \\
1.5 ~\dex < \log {\it g} < 3.5~\dex  , \\
-1.0 ~\dex < [Fe/H] < 0.5 ~\dex  , \\
$single-lined spectra only.$
\end{cases}$

\vspace*{3mm}

Furthermore, the selected sources have at least two-night
exposures in LAMOST database, and the largest radial
velocity variation $\Delta V_{\rm R} \ga 70~\km/s$.
Consequently, we obtain a sample of 43 single-lined
binaries. In addition, the sources without {\it Gaia}
parallax or with negative parallax have been removed
\citep{Jayasinghe2019, Ziaali2019}. We crossmatch the
sources with the ASAS-SN Sky
Patrol\footnote{\url{https://asas-sn.osu.edu}} database
\citep{Kochanek2017, Shappee2014}, and therefore we
derive a sample of 17 binaries with periods longer
than 5 days (the reason is given in the fourth
paragraph of Section~\ref{sec3.1}). We refer to the
$\rm{S/N}$ in Equation (6) from \citet{Hartman&Bakos(2016)}
to measure the significance for peaks identified in
the periodogram.
Finally, we obtain 9 sources with
$\rm{S/N}(\it{T}\rm_{ph}) > 30$ as our sample, which
are shown in Table~\ref{T1}.
We also crossmatch our sample with simbad in $5^{\arcsec}$,
and find that only Source number 3 has X-ray
information (refer to Section~\ref{sec3.1}).

Since the sources in our sample are all with
single-lined spectra, the unseen object in a binary is
therefore either a compact object or a much fainter star,
roughly speaking, less than 10\% of the luminosity of the
observed giant star. The luminosity is shown in column
14 of Table~\ref{T1}, which is calculated by the
apparent magnitude from UCAC4 and the parallax from
{\it Gaia} DR2, where the bolometric correction and
extinction have been taken into consideration. It is seen
from Table~\ref{T1} that the luminosity of these sources
is less than 100 solar luminosities. If the unseen object
is a main sequence star or a subgiant star with 3 solar
masses, it will be more than 30 solar luminosities.
Thus, it ought to be observed and the corresponding
spectra of the binary should not be the single-lined type.
Thus, once $M_2 > 3M_{\sun}$ is matched, the unseen
object has high possibility to be a BH. In this work,
we manage to search for BH candidates following this spirit.
We would point out another possibility that the
system is a triple system. For example, if the system consists of
a giant star and a pair of 1.5 solar mass stars in a close binary,
then the total luminosity of the pair of stars
is around 10 solar luminosities, and therefore may be optically invisible.

\section{Results}\label{sec3}

\subsection{Period analyses}\label{sec3.1}

\begin{figure}[htb!]
\includegraphics[width=20cm,height=18cm]{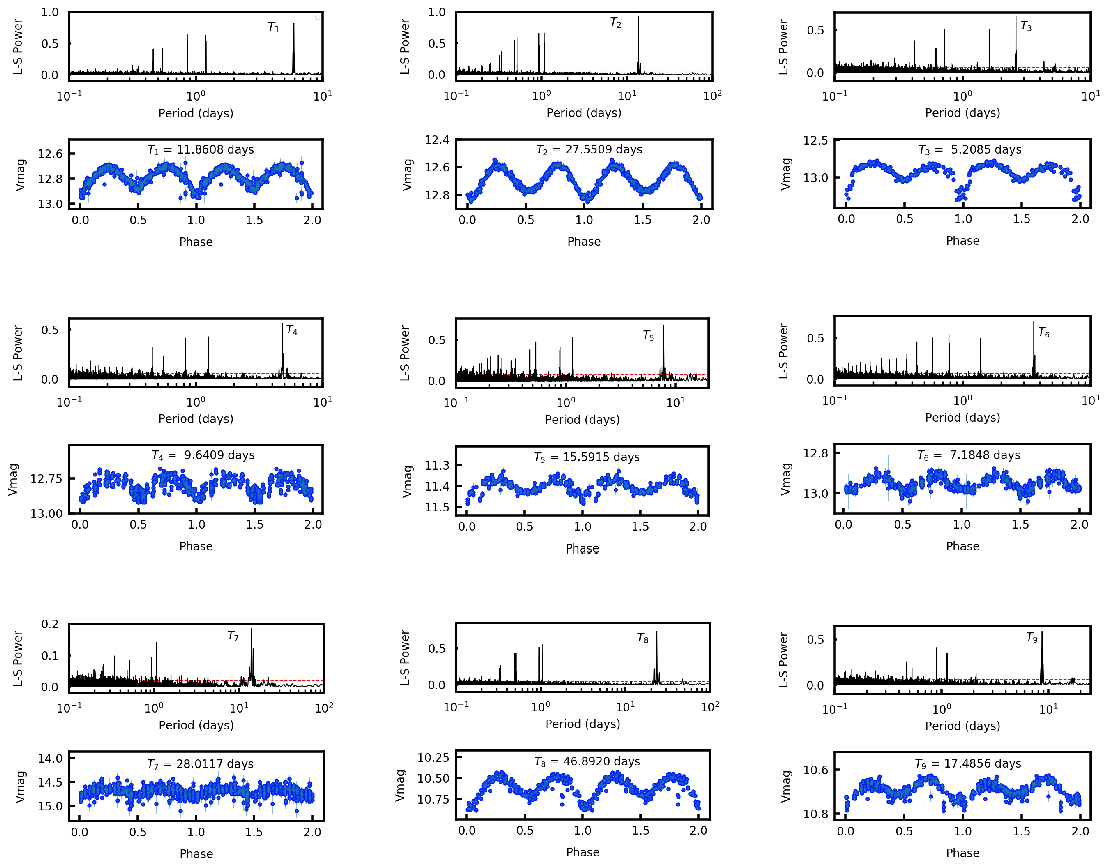}
\setlength{\abovecaptionskip}{-3cm}
\caption{Light curves of the nine sources in Table~\ref{T1}
folded by the Lomb-Scargle algorithm, where the period of variability
is shown in each panel.
\label{F1}}
\end{figure}

The Lomb-Scargle method \citep{Lomb1976, Scargle1981}
is a useful statistical tool to extract periodic
signals in unevenly-spaced data. This method can
reflect the intensity of the captured periodic signal
on the power peaks. The Lomb-Scargle periodogram is
calculated, the period corresponding to the highest
power was extracted. Then light curves can be folded
with the retrieved periods. We searched for the
periods for the sources in our sample, and five of
them (Sources number 1-3, 7 and 8) have been investigated
by \citet{Gu2019} without any information of period.
The folded light curves for the nine sources are shown
in Figure~\ref{F1}.
As mentioned in Section~\ref{sec2},
only Source number 3 was known as a faint X-ray source
according to the {\it ROSAT} observations \citep{Voges2000}.
The light curve of this source in Figure~\ref{F1} shows that
it may have two possibilities.
One is an eclipsing binary caused by an accretion disk,
which coincides with the X-ray observations.
The other possibility is an eclipsing binary of the Algol
type where the X-ray emission is related to an active star.
From the shape of light curves, the other eight sources
in our sample may be either the ellipsoidal modulation
type or the eclipsing binary type.
We would stress that, for both of these two mechanisms,
the periodic variability can reveal the orbital period.
Thus, we can evaluate the mass of the unseen object
by using the orbital period.

We compare our derived period based on the Lomb-Scargle
algorithm with that given by ASAS-SN. We found that our
results are identical to that from the ASAS-SN website,
except for Source number 2. The photometric period
(55.1046 days) from the ASAS-SN website for this source
is exactly twice of ours (27.5509 days). In our opinion,
the different periods may result from different folding
algorithms. For this source, we adopt our period in the
following analyses.

Even though the shape of folded light curves may indicate
that the photometric period $T_{\rm ph}$ is identical
with the orbital period $P_{\rm orb}$, some analyses are
required to confirm that. In a binary, the relation between
the separation $a$ and $P_{\rm orb}$ takes the form:
\begin{equation}
\frac{G(M_1+M_2)}{a^3} = \frac{4\pi^2}{P_{\rm orb}^2} \ .
\label{E1}
\end{equation}
In addition, the Roche-lobe radius of the optically
visible star $R_{\rm L1}$
can be expressed as \citep{Paczynski1971}:
\begin{equation}
\frac{R_{\rm L1}}{a} = 0.462 \left( \frac{M_1}{M_1 + M_2} \right) ^{1/3} \ .
\label{E2}
\end{equation}
Based on the reasonable assumption that the radius
of the giant star is no larger than the Roche-lobe
radius, i.e., $R_1 \leqslant R_{\rm L1}$ \citep{Gu2019},
and by combining Equations~(\ref{E1}) and~(\ref{E2}),
the following inequality can be derived:
\begin{equation}
P_{\rm orb} \geqslant 2\pi \left[ \frac{(R_1/0.462)^3} {G M_1} \right]^{1/2} \ .
\label{E3}
\end{equation}
Thus, there exists a lower limit for the orbital period
once $R_1$ and $M_1$ (or simply the mass density $\rho_1$)
is derived:
\begin{equation}
P_{\rm orb}^{\rm min} = 0.369 ~(\rho_1/\rho_{\sun})^{-1/2}
~{\rm days} \ ,
\label{E4}
\end{equation}
where $\rho_{\sun}$ is the solar density.

As indicated by Equation~(\ref{E4}), if the optically
visible star is of main sequence, the orbital period
can be less than one day, which well agrees with most
confirm BHs in low-mass X-ray binaries. In this work,
however, we focus on the cases that the companion is
a late-type giant star, such as a red giant. For instance,
given $M_1 = M_{\sun}$ and $R_1 = 10 R_{\sun}$,
Equation~(\ref{E4}) results in $P_{\rm orb}^{\rm min}
\approx 11.7$~days. That is why we have focused on the
nine sources with $T_{\rm ph} > 5$~days. Otherwise,
the photometric period $T_{\rm ph}$ is unlikely the
orbital period $P_{\rm orb}$.

\begin{figure}
\plotone{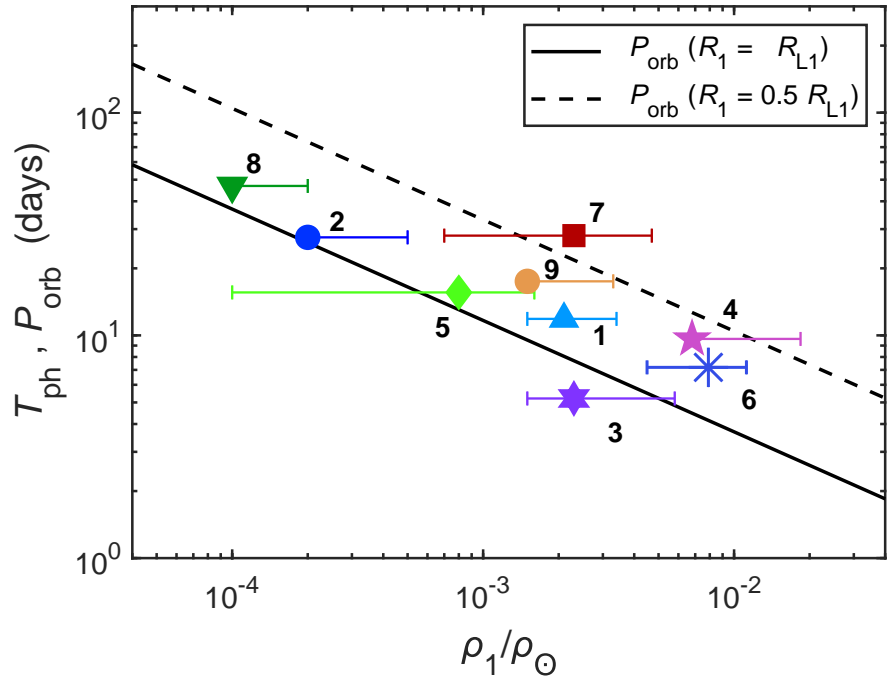}
\caption{
A comparison of the analytic orbital period $P_{\rm orb}$
(lines) with the observational variability period
$T_{\rm ph}$ (symbols). The solid line represents the lower
limit $P_{\rm orb}^{\rm min}$ calculated by Equation~(\ref{E4}), where the giant star fills its Roche lobe
($R_1 = R_{\rm L1}$).
The dashed line corresponds to a case that the Roche lobe
is not filled, with $R_1 = 0.5 R_{\rm L1}$.
\label{F2}}
\end{figure}

For the sources in our sample, apparent periodic
variability ($0.1\sim 0.5$~mag) has been observed.
If the periodic variability is related to the ellipsoidal modulation, then the radius of optically visible star
cannot be far below the corresponding Roche-lobe radius.
Consequently, the period $T_{\rm ph}$ should not be far
beyond $P_{\rm orb}^{\rm min}$. Figure~\ref{F2}
shows the consistency of the photometric period and
the orbital period.
A comparison of the observations with our analyses
is shown in Figure~\ref{F2}, where the analytic
$P_{\rm orb}^{\rm min}$ for $R_1 = R_{\rm L1}$ (solid line)
is calculated by Equation~(\ref{E4}).
In addition, since the giant star may not fill
its Roche lobe, we also plot the analytic $P_{\rm orb}$
for $R_1 = 0.5 R_{\rm L1}$ (dashed line).
The observations for the nine sources are denoted by
different symbols, where $M_1$ and $R_1$ are derived
by the stellar evolution model, as mentioned in
Table~\ref{T1}. It is seen from Figure~\ref{F2} that
all the nine sources are well located around the solid
line or between the solid line and the dashed line,
which indicates that the relation $P_{\rm orb} = T_{\rm ph}$
is quite reasonable. As a consequence, we can place
better constraints on the mass $M_2$ with the values
of $P_{\rm orb}$.

\subsection{Mass measurement}\label{sec3.2}

We evaluate the mass $M_1$ of the nine sources in
our sample by the PARSEC model \footnote{\url{http://stev.oapd.inaf.it/cgi-bin/cmd_3.1}}
\citep{Bressan2012, Marigo2008},
and measure the mass of optically invisible companion
by the equation of mass function. The well-known mass
function for $M_2$ takes the form \citep{RM2006}:
\begin{equation}
f(M_2) \equiv \frac{M_2  \sin^3 i}{(1+q)^2} =
\frac{K_1^3 P_{\rm orb}}{2 \pi G}
\ ,
\label{E5}
\end{equation}
where $K_1 \geqslant \Delta V_{\rm R} /2$ is the
semi-amplitude of the giant star, and the
mass ratio is defined as $q\equiv M_1/M_2$.
In the above equation, once $K_1$ and $P_{\rm orb}$
are given, the mass function $f(M_2)$ can be obtained,
and it is certain that $M_2 > f(M_2)$.
In addition, if $M_1$ can be derived from the spectra
and $\sin i$ is provided, then $M_2$ can be well
constrained. Referring to the inclination angle of
most BH binaries in \citet{Corral-Santana2016},
we assume a typical inclination angle $i=60^\circ$
in this work.

As shown in Table~\ref{T1}, $M_1$ of most sources are
in the range of $1 M_\odot < M_1 < 2 M_\odot$.
If $M_2 > 3 M_{\sun}$ can be matched, the source can
be regarded as a BH candidate. However, it is not easy
to directly satisfy such a condition. On the other
hand, if $M_2 > M_1$ can be matched, according to the
rule of stellar evolution, the optically invisible
star is likely to be a compact object (except for the
Algol case, see below).
Thus, we plot four theoretical lines in Figure~\ref{F3}
(from bottom to top) corresponding to
($M_1 = 1 M_\odot$, $M_2 = M_1$),
($M_1 = 2 M_\odot$, $M_2 = M_1$),
($M_1 = 2 M_\odot$, $M_2 = 3 M_\odot$), and
($M_1 = 1 M_\odot$, $M_2 = 3 M_\odot$), respectively.

It is seen from Figure~\ref{F3} that, there is no
source above or even in the upper shaded region,
which means that there is no strong BH candidate
according to the current observations. On the other hand,
Source number 8 is located well above the green
shaded region, and close to the blue shaded region.
Even under the extreme case with the inclination angle $i=90^\circ$, we also derive $M_2 > M_1$.
Hence Source number 8 is probably a compact object.
Whether or not it is a BH requires follow-up
spectroscopic observations to obtain the radial
velocity curve. However, we would point out that there
exist some binaries like Algol, in which the lower mass
star is more ``evolved" than its companion. Thus,
mass exchange can allow $M_2 > M_1$ with the
more massive star being on the main sequence. In other words,
the condition $M_2 > M_1$ may imply a compact star
but is not a sufficient condition.
In addition, most sources in Table~\ref{T1} have only two or
three observations, the semi-amplitude of radial
velocity $K_1$ may be significantly larger than the
current $\Delta V_{\rm R}/2$. Thus, it is quite
possible for the mass $M_2$ to be significantly higher
than the current evaluation. We therefore use the
black arrows in Figure~\ref{F3} to show such an increase possibility.

\section{Conclusions and discussion}\label{sec4}

In this work, we have proposed the method to search
for stellar-mass BH candidates by including the
LAMOST spectra and the ASAS-SN photometry, where
the orbital period $P_{\rm orb}$ may be revealed
by the periodic light curve. We have obtained a
sample of 9 single-lined spectroscopic binaries
containing a giant star with large radial velocity
variation $\Delta V_{\rm R} \ga 70~\km/s$, and the
photometric period of the sources are satisfy
$T_{\rm ph} > 5$~days. Moreover, based on the
relation $R_1 \leqslant R_{\rm L1}$, we have
checked that $T_{\rm ph}$ and $P_{\rm orb}$ are
likely identical for the sources in our sample.
As a consequence, the mass $M_2$ can be better
constrained. We have shown that Source number 8
is likely to be a compact object. It is worth
follow-up spectroscopic observations to check
whether it is a BH. Moreover, for the other sources,
the real mass $M_2$ can be significantly higher
than the current evaluation. Thus, they are also
potential BH candidates. In our opinion, it is an
efficient method to constrain $M_2$ by combining
the LAMOST spectra and the ASAS-SN photometry.

In this work, we have focused on the giant companion.
In fact, our method is also valid for the main sequence
star case. Normally, the orbital period $P_{\rm orb}$
of a main sequence star is significantly shorter
(less than one day, as implied by Equation~(\ref{E4}))
for the ellipsoidal modulation type. In such case,
the radial velocity variation in the same night can
provide crucial information. Thus, once the single
exposure spectra of LAMOST are released, many more
BH candidates can be found through our method.
On the other hand, the LAMOST Medium Resolution
Survey will provide more accurate radial velocity
and more repeating exposures (around 60 exposures
for a source in the time-domain spectroscopic survey),
which enable us to derive a clear radial velocity
curve and make better constraint on the mass of candidates.

The sources in our sample are binaries with relatively long orbital
periods ($5\sim 47$~days). However, the {\it Gaia} DR2 solution
has assumed a single star model
and has mistaken the binary motion itself as part of the parallax, which
may result in systematic errors for the parallax and distance.
Whether the real parallaxes are larger or smaller than the current values
is related to the observational times. For the cases with adequate
observations by {\it Gaia}, the real parallaxes will be smaller.
On the contrary, for the cases with inadequate observations,
the results will be quite uncertain.
In DR3, non-single star model will be considered in data analysis.
In full release for the nominal mission, the catalog will provide
all available variable-star and non-single-star solutions.

\acknowledgments

We thank Mou-Yuan Sun, Wei-Kai Zong, Xuefei Chen, Zhaoxiang Qi, and
Kento Masuda for helpful discussions, and the referee
for constructive suggestions that improved the paper.
This work was supported by the National Natural Science
Foundation of China (NSFC) under grants 11573023, 11603035, 11603038,
U1831205 and 11425313, as well as was developed
in part at the 2018 Gaia-LAMOST Sprint workshop,
supported by the National Natural Science Foundation of
China (NSFC) under grants 11333003 and 11390372,
and the Fundamental Research Funds for the Central
Universities under grants 20720190122, 20720190115, and 20720190051.
This work has made use of data products from
the Guoshoujing Telescope (the Large Sky Area Multi-Object
Fiber Spectroscopic Telescope, LAMOST)
and the All-Sky Automated Survey for Supernovae (ASAS-SN).
LAMOST is a National Major Scientific Project built by the
Chinese Academy of Sciences.
Funding for the project has been provided by the National Development
and Reform Commission. LAMOST is operated
and managed by the National Astronomical Observatories,
Chinese Academy of Sciences. ASAS-SN is hosted by Las Cumbres Observatory,
we thank the Las Cumbres Observatory and
its staff for its continuing support of the ASAS-SN project.

\clearpage

\clearpage

\clearpage

\begin{figure}
\plotone{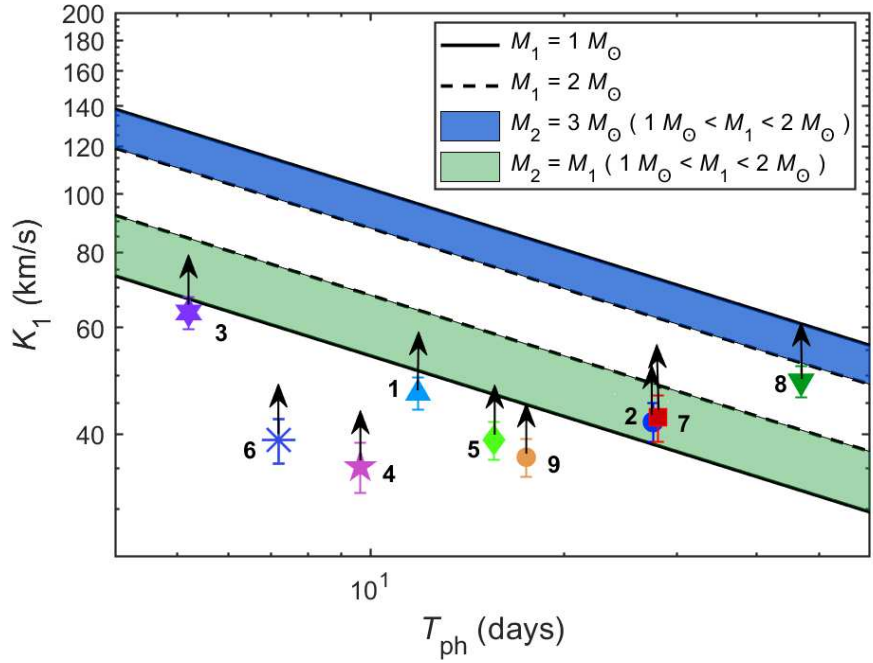}
\caption{
A comparison between analyses and observations in the
$K_1-T_{\rm ph}$ diagram, where the semi-amplitude $K_1$
is no less than half of the observed maximal
variation in a few repeating observations
($K_1 \geqslant \Delta V_{\rm R} /2$).
\label{F3}}
\end{figure}

\clearpage

\movetabledown=0.2in
\begin{longrotatetable}
\begin{deluxetable}{lcccccccccccccccc}
\tablecaption{Parameters for the sources in our sample.
\label{T1}}
\tablewidth{1000pt}
\tablehead{
\colhead{No.}
&\colhead{R.A.}
&\colhead{Decl.}
&\colhead{$T_{\rm ph}$}
&\colhead{$\rm{S/N}(\it{T}\rm_{ph})$}
&\colhead{$T_{\eff,L}$}
& \colhead{log \it{g}}
& \colhead{$\rm [Fe/H]$}
&\colhead{$N_{\rm obs}$}
& \colhead{$\Delta V_{\rm R}$}
&\colhead{$\varpi$}
&\colhead{Vmag}
&\colhead{Kmag}
&\colhead{$L$}
&\colhead{$R_{1}$}
& \colhead{$M_{1}$}
& \colhead{$M_{2}$}
\\
\colhead{}& \colhead{}& \colhead{}& \colhead{(days)}
& \colhead{}& \colhead{($\K$)}& \colhead{($\dex$)}& \colhead{($\dex$)}&\colhead{}& \colhead{($\km/s$)}
& \colhead{($\rm mas$)}
& \colhead{(mag)}& \colhead{(mag)}& \colhead{($L_{\sun}$)}
& \colhead{($R_{\sun}$)}& \colhead{($M_{\sun}$)}& \colhead{($M_{\sun}$)}\\
\colhead{(1)}& \colhead{(2)}& \colhead{(3)}& \colhead{(4)}& \colhead{(5)}& \colhead{(6)}& \colhead{(7)}& \colhead{(8)}& \colhead{(9)}& \colhead{(10)}&\colhead{(11)}&\colhead{(12)}& \colhead{(13)}&\colhead{(14)}&\colhead{(15)}&\colhead{(16)}&
\colhead{(17)}
}
\centering
\tablewidth{0pt}
\tabletypesize{\tiny}
\tablecolumns{17}
\tablenum{1}
\startdata
1	&	0.839201575	&	38.51855052	&	11.8608 	&	118.56 	
&	4696 	$\pm$	57 	&	2.65 	$\pm$	0.09 	&	-0.25 	$\pm$	0.05 	&	3	&	93.5 	$\pm$	5.6 	&	0.505 	$\pm$	0.043 	&	12.736 	&	9.952 	&	50.176 	&
$	8.1	^{ +	1.5	}_{ -	0.8	}$&$	1.1	^{+	0.3	}_{-	0.1	}$&$	0.9 	^{+	0.2 	}_{-	0.1 	}$	\\
2$^{{\dag}}$	&	3.887105349	&	38.68886824	&	27.5509 	
&	116.33 	&	4301 	$\pm$	44 	&	1.95 	$\pm$	0.07 	
&	-0.47 	$\pm$	0.04 	&	4	&	83.7 	$\pm$	6.2 	
&	0.379 	$\pm$	0.032 	&	12.665 	&	9.641 	&	91.627 	
&$	17.8	^{ +	0.6	}_{ 		}$&$	0.9	^{+	1.5	}_{		}$&$	1.1 	^{+	0.9 	}_{		}$	\\
3	&	74.05325351	&	54.00589535	&	5.2085 	&	55.65 	&	4769 	$\pm$	106 	&	2.68 	$\pm$	0.17 	&	-0.31 	$\pm$	0.10 	&	6	&	127.2 	$\pm$	7.8 	&	1.068 	$\pm$	0.034 	&	12.784 	&	9.004 	&	31.245 	&
$	7.8 	^{ +	3.5	}_{ -	0.7	}$&
$	1.1 	^{+	0.7	}_{-	0.2	}$&
$	1.0 	^{+	0.4 	}_{-	0.2 	}$	\\
4$^{{\dag}}$	&	82.31076394	&	42.09587217	&	9.6409 	&	39.49 	&	4700 	$\pm$	98 	&	2.97 	$\pm$	0.15 	&	-0.36 	$\pm$	0.09 	&	4	&	70.5 	$\pm$	6.7 	&	1.398 	$\pm$	0.048 	&	12.727 	&	9.941 	&	6.774 	&
$	5.1 	^{ +	2.8	}_{ 		}$&$	0.9 	^{+	0.4	}_{		}$&$	0.5 	^{+	0.2 	}_{		}$	\\
5	&	93.81977494	&	22.11031808	&	15.5915 	&	32.38 	
&	4796 	$\pm$	25 	&	2.43 	$\pm$	0.04 	&	-0.47 	$\pm$	0.02 	&	2	&	78.1 	$\pm$	5.6 	&	1.090 	$\pm$	0.072 	&	11.362 	&	8.051 	&	77.445 	&$	12.0 	^{ +	3.3	}_{ -	3.4	}$&$	1.4 	^{+	0.6	}_{-	0.5	}$&$	0.9 	^{+	0.3 	}_{-	0.2 	}$	\\
6	&	102.0930387	&	21.82487008	&	7.1848 	&	37.92 	
&	5093 	$\pm$	29 	&	3.11 	$\pm$	0.05 	&	-0.27 	$\pm$	0.03 	&	2	&	78.1 	$\pm$	6.6 	&	0.539 	$\pm$	0.042 	&	12.920 	&	10.231 	&	42.086 	&$	6.0 	^{ +	0.8	}_{ -	0.8	}$&$	1.7 	^{+	0.2	}_{-	0.3	}$&$	0.7 	^{+	0.2 	}_{-	0.1 	}$	\\
7	&	111.3363737	&	28.06745981	&	28.0117 	&	42.72 	
&	4833 	$\pm$	188 	&	2.75 	$\pm$	0.30 	&	-0.23 	$\pm$	-0.23 	&	6	&	85.1 	$\pm$	7.6 	&	0.152 	$\pm$	0.032 	&	14.698 	&	12.075 	&	82.412 	&$	8.7 	^{ +	2.8	}_{ -	1.9	}$&$	1.5 	^{+	0.6	}_{-	0.4	}$&$	1.4 	^{+	0.5 	}_{-	0.3 	}$	\\
8$^{{\dag}}$	&	169.1246518	&	55.72840217	&	46.8920 	
&	74.31 	&	4191 	$\pm$	102 	&	1.82 	$\pm$	0.16 	&	-0.75 	$\pm$	0.10 	&	3	&	97.8 	$\pm$	5.8 	
&	1.086 	$\pm$	0.031 	&	10.638 	&	7.377 	&	82.415 	
&$	20.8	^{ +	8.1	}_{ 		}$&$	0.9	^{+	0.5	}_{		}$&$	1.9 	^{+	0.6 	}_{		}$	\\
9	&	325.3386324	&	28.4225968	&	17.4856 	&	40.81 	
&	4770 	$\pm$	78 	&	2.51 	$\pm$	0.12 	&	-0.15 	$\pm$	0.07 	&	2	&	73.2 	$\pm$	5.2 	&	1.063 	$\pm$	0.036 	&	10.675 	&	8.036 	&	67.864 	&$	10.8 	^{ +	4.1	}_{ -	0.2	}$&$	1.9 	^{+	0.6	}_{-	0.7	}$&$	1.1 	^{+	0.3 	}_{-	0.4 	}$	\\
\enddata
\tablecomments{$^{{\dag}}$ The lower limit of
$M_1$ or $R_1$ cannot be well estimated from the PARSEC model.
Column (1): number of the source.
Column (2): R.A. (J2000).
Column (3): decl. (J2000).
Column (4): folded period from the ASAS-SN photometry.
Column (5): significance of the periodogram.
Column (6): effective temperature from LAMOST.
Column (7): surface gravity from LAMOST.
Column (8): metallicity from LAMOST.
Column (9): times of observations.
Column (10): observed largest variation of radial velocity.
Column (11): parallax from {\it Gaia}.
Column (12): V-band magnitude from UCAC4.
Column (13): K-band magnitude from UCAC4.
Column (14): luminosity calculated by the
apparent magnitude from UCAC4 and the parallax from
{\it Gaia} DR2.
Column (15): radius of the giant star from the PARSEC model.
Column (16): mass of the giant star from the PARSEC model.
Column (17): mass of the invisible star for
``$i = 60^\circ$, $K_1 = \Delta V_{\rm R}/2$,
and $P_{\rm orb} = T_{\rm ph}$".}
\end{deluxetable}
\end{longrotatetable}

\clearpage

\end{document}